\documentclass[9pt,twocolumn,twoside]{opticajnl}
\journal{opticajournal} 

\setboolean{shortarticle}{true}

\usepackage{lineno}
\usepackage{braket}
\usepackage{changes}

\title{Engineering walk-off-induced orbital angular momentum spectrum in spontaneous parametric downconversion}

\author[1,*]{Yang Xu}

\author[1,2,3]{Robert W. Boyd}

\affil[1]{Department of Physics and Astronomy, University of Rochester, Rochester, New York 14627, USA}
\affil[2]{The Institute of Optics, University of Rochester, Rochester, New York 14627, USA}
\affil[3]{Department of Physics, University of Ottawa, Ottawa, Ontario K1N 6N5, Canada}

\affil[*]{yxu100@ur.rochester.edu}

\begin{abstract}
Spontaneous parametric downconversion (SPDC) has been considered as a reliable source of high-dimensional entangled states in orbital angular momentum (OAM) basis. In real-world experiments, the spatial walk-off of the pump often degrades the fidelity of the generated quantum state. Since the walk-off effect breaks the rotational symmetry of the system, the conservation of total OAM is violated. Although the compensation of walk-off effects has become a well-established experimental technique, a systematic modal analysis of the spatial walk-off effect is still incomplete for SPDC. Here, we quantitatively analyze the violation of OAM conservation due to the pump walk-off effect in SPDC processes. We have derived a scaling law of the total OAM distribution with respect to the pump walk-off angle. We have also explored the feasibility of using the spatial walk-off as a mechanism to engineer the generated quantum state. Our study has provided guidelines for the generation of OAM-entangled state under realistic experimental conditions.
\end{abstract}

\setboolean{displaycopyright}{false} 

\begin{document}

\maketitle

The spatial modes of electromagnetic waves, particularly Laguerre-Gauss (LG) modes with an azimuthal phase $e^{il\phi}$, carry a well-defined orbital angular momentum (OAM) of $l\hbar$ where $l$ takes the integer value. OAM modes constitute a complete set of orthogonal spatial modes that defines a discrete high-dimensional Hilbert space \cite{yan2014high, willner2017recent, PhysRevResearch.6.L042047}. This dimensionality is highly advantageous for many classical and quantum applications such as information encoding and secure communication protocols. For example, in OAM-based satellite quantum key distribution (QKD) \cite{wang2020satellite, Halevi:24}, the protocol depends on angle-OAM entanglement because the high dimensionality allows more information capacity and better noise resilience \cite{willner2021orbital}.

Spontaneous parametric downconversion (SPDC) in a $\chi^{(2)}$ crystal is one of the most reliable and popular methods to produce these OAM-entangled photon pairs. The use of high-dimensional entanglement produced by SPDC has led to innovative applications including the teleportation of high-dimensional quantum states and the generation of multiphoton entanglement \cite{erhard2020advances}. Recently, the conservation of OAM on a single photon level has also been demonstrated in the laboratory \cite{PhysRevLett.134.203601}, opening new avenues for the direct generation of multiphoton high-dimensional entanglement in free space. Alongside these experimental efforts, a large number of theoretical works have been dedicated to the characterization of the two-photon quantum state in LG basis \cite{mair2001entanglement, quesada2015time, d2021full, PhysRevResearch.4.033098}. However, most theoretical models of SPDC often assume rotational symmetry and thus downplay the effect of spatial walk-off. This approximation remains valid only under the thin-crystal limit. As the demand for bright entangled sources has popularized the use of long nonlinear crystals and tightly focused pump beams, the broken rotational symmetry caused by the spatial walk-off of the pump becomes pronounced, violating the conservation of total OAM \cite{torres2003quantum, torres2005spatial}. Although experimental techniques to compensate for walk-off effects \cite{kanseri2020effect, lee2021sagnac, PhysRevA.110.063515} are well-established, a systematic analysis of spatial walk-off in the generation of OAM-entangled photon pairs is still inadequate.

In this Letter, we address the gap in the current theoretical research on SPDC by providing a numerical study of the effect of the pump walk-off on the OAM spectrum of the downconverted two-photon state. We quantitatively analyze the violation of OAM conservation of the optical subsystem \cite{PhysRevLett.130.153803} due to the pump walk-off effect in type-I SPDC processes. By solving the evolution of the two-photon state with perturbation theory and OAM mode decomposition, we have thoroughly explored the role of spatial walk-off in SPDC and have proposed practical methods that utilize the pump walk-off as a tuning knob for OAM-entangled state engineering. Our model is fully compatible with type-0 quasi-phase-matching and can be adapted to type-II phase matching with modest modifications (Supplement 1). 

\begin{figure}[hbt!]
\centering
\hspace*{-0.4cm}
\includegraphics[width=0.37\textwidth]{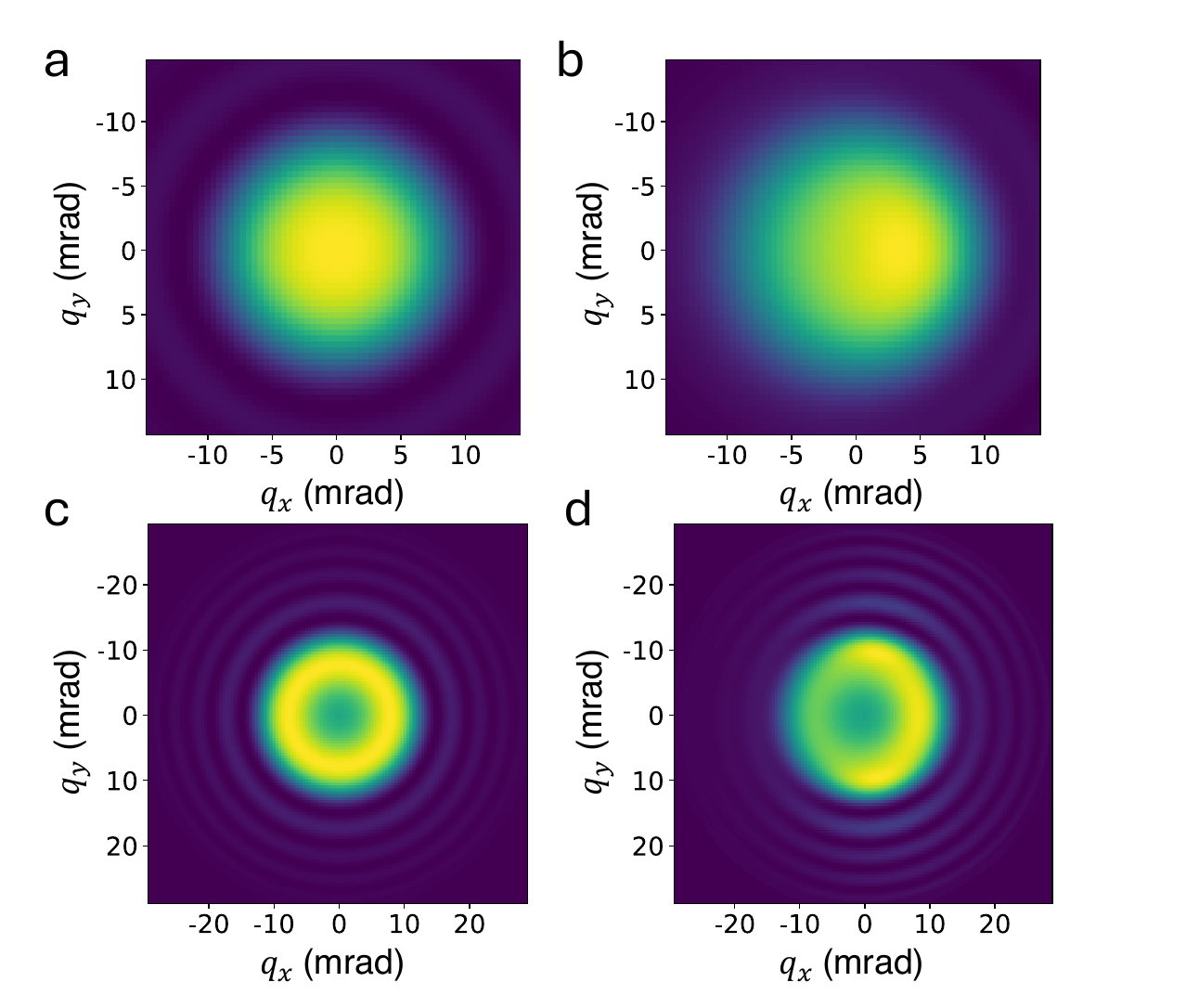}
\caption{Far-field intensity profiles of generated signal field at 710 nm. The type-I SPDC is driven by a Gaussian pump field with a waist of 0.2 mm at 355 nm. The direction of the walk-off is assumed to be along the $x$-axis. \textbf{a.} Intensity profile of downconverted signal at collinear phase matching with no pump walk-off ($\rho = 0$). \textbf{b.} Intensity profile of downconverted signal at collinear phase matching with a pump spatial walk-off of $\rho = 3^{\circ}$. \textbf{c.} Intensity profile of downconverted signal at non-collinear phase matching with no pump spatial walk-off ($\rho = 0$). \textbf{d.} Intensity profile of downconverted signal at non-collinear phase matching with a spatial walk-off of $\rho = 3^{\circ}$.  }
\label{fig:fig1}
\end{figure}

For an SPDC process, the Hamiltonian in the interaction picture takes the following form:
\begin{align}
    \mathcal{H} = i\hbar\Gamma \int d^3\mathbf{q}_i d^3\mathbf{q}_s \Phi(\mathbf{q}_s, \mathbf{q}_i)  \hat{a}_{\mathbf{q}_s}^\dagger \hat{a}_{\mathbf{q}_i}^\dagger +\text{h.c.} \label{eq:hamilt}
\end{align}
where the subscripts $p, s, i$ stand for the pump, signal, and idler, respectively. $\Phi(\mathbf{q}_s, \mathbf{q}_i)$ is the two-photon wavefunction
\begin{align}
    \Phi(\mathbf{q}_s, \mathbf{q}_i) = \Phi(q_s, q_i, \phi_s, \phi_i) =C \tilde{E_p}(\mathbf{q}_s + \mathbf{q}_i) \text{sinc}\left(\frac{\Delta k L}{2}\right)e^{\frac{i\Delta k L}{2}}\label{eq:two-photon}.
\end{align}
Here, $\Gamma$ is the nonlinear coupling constant, $\mathbf{q}_{s,i}$ is the transverse component of the wavevectors for signal and idler, $q_s, q_i$ are the magnitude of the transverse wavevectors of the signal and idler, $\phi_s, \phi_i$ are the azimuthal angles, $\hat{a}_{\mathbf{q}_{s,i}}^\dagger$ is the creation operator for the corresponding plane-wave mode, $C$ is the normalization constant, $\tilde{E_p}(\mathbf{q}_s + \mathbf{q}_i)$ is the Fourier-transformed transverse spatial profile of the pump, $L$ is the length of the nonlinear downconversion crystal, and $\Delta k = k_{pz} - k_{sz} - k_{iz}$ is the phase mismatch along the $z$-direction in which the pump is propagating. 

\begin{figure}[hbt!]
\centering
\hspace*{-0.6cm}
\includegraphics[width=0.5\textwidth]{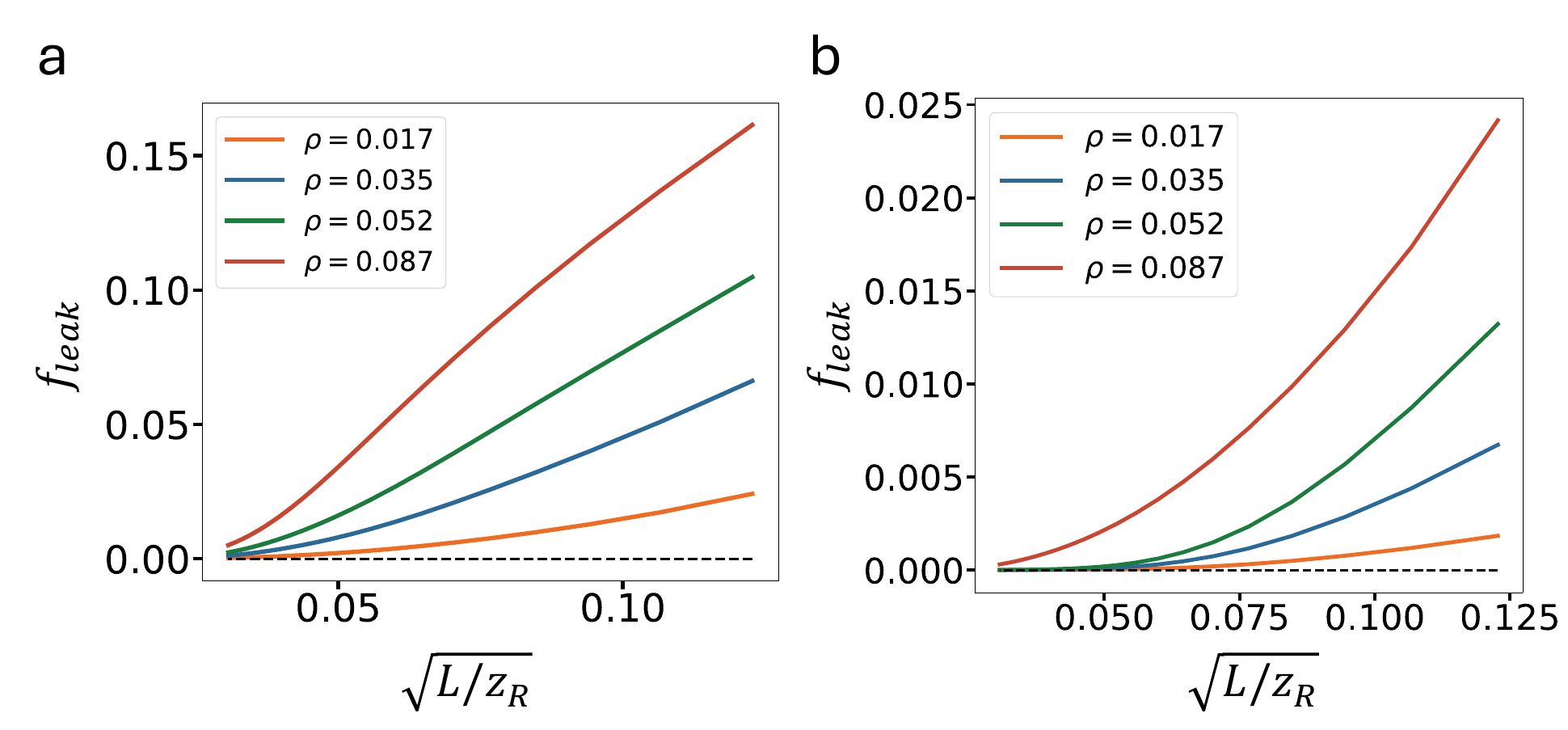}
\caption{Numerical dependence of $f_{leak}$ on the focusing parameter of the pump beam $\sqrt{L/z_R}$ for multiple pump walk-off angles under \textbf{a.} collinear phase matching condition and \textbf{b.} non-collinear phase matching condition. The probability of measuring a change in total OAM (i.e., violation of OAM conservation) becomes more pronounced for long crystals and a focused pump. Furthermore, we observe that the non-collinear geometry exhibits greater resilience to the walk-off effect. }
\label{fig:fig2}
\end{figure}

\begin{figure*}[hbt!]
\centering
\includegraphics[width=0.65\textwidth]{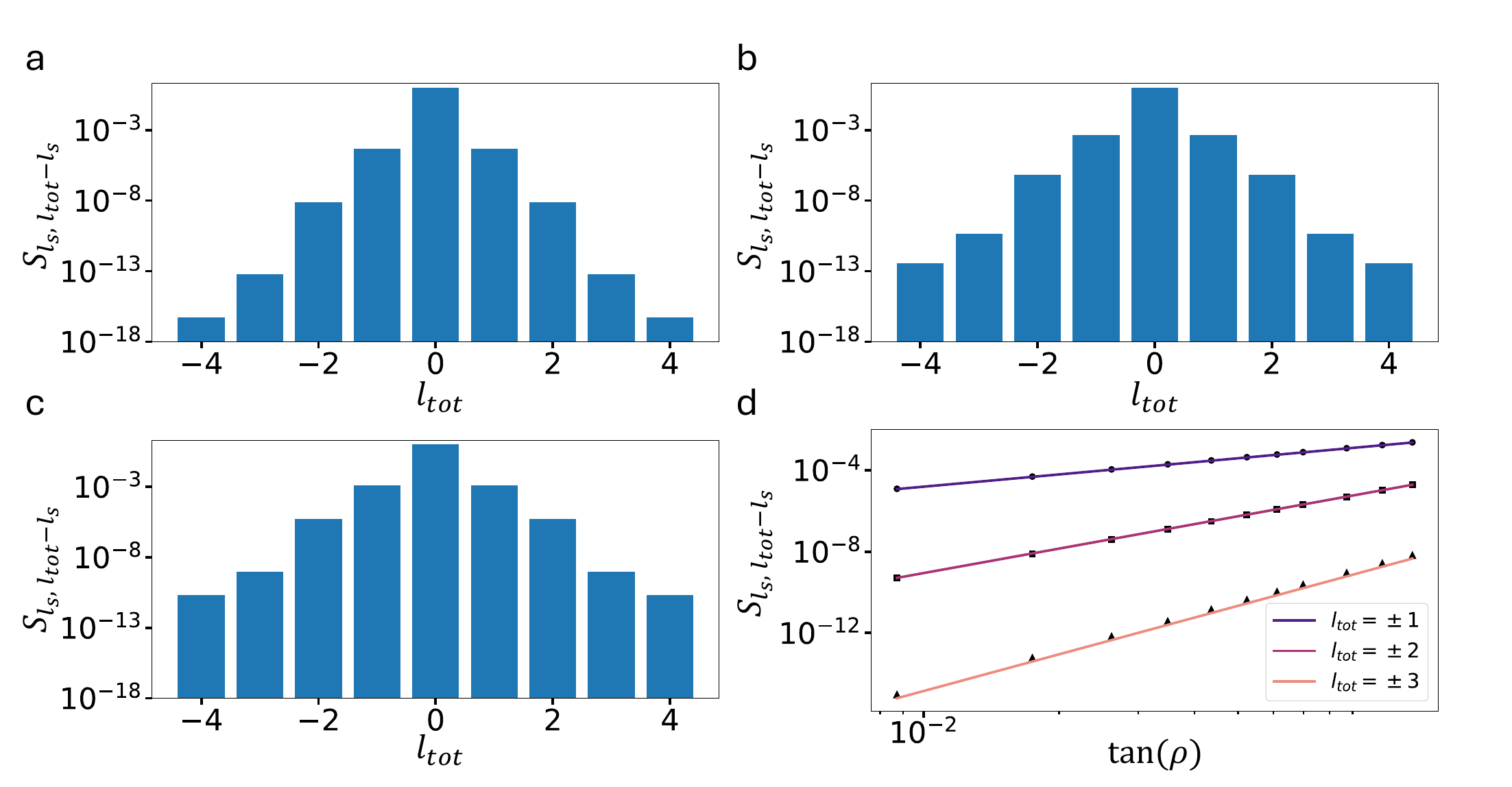}
\caption{Total OAM distribution and its scaling with the pump spatial walk-off. The type-I SPDC is pumped by a Gaussian beam with $l_{tot} = l_p = 0$. \textbf{a-c.} Probability distribution of the total OAM for a walk-off angle of \textbf{a.} 1$^\circ$, \textbf{b.} 3$^\circ$ and \textbf{c.} 5$^\circ$ under the collinear phase matching condition. Note that the coupling to non-zero OAM modes becomes significant at larger walk-off angles, suggesting a stronger effect of broken rotational symmetry. \textbf{d.} Scaling of the first three total OAM sidebands ($\left|l_{tot}\right| = 1, 2,3$) under the small walk-off approximation. The probability of having a change of $m\hbar$ in the total OAM scales with $m$-th order azimuthal harmonic in the Jacobi-Anger expansion: $S_{l_s,\text{ } m-l_s} \propto (\text{tan} \rho)^{2\left|m\right|} \approx \rho^{2\left|m\right|}$. The purple, maroon and the orange lines show the polynomial dependence of $(\text{tan} \rho)^2$, $(\text{tan} \rho)^4$ and $(\text{tan} \rho)^6$ respectively.}
\label{fig:fig3}
\end{figure*}

For simplicity, we focus our discussion only on type-I degenerate SPDC with a monochromatic pump where $\omega_p = 2\omega_s = 2\omega_i$. Then, the OAM spectrum $S_{l_s,l_i}$ of the two-photon state can be determined by \cite{PhysRevResearch.4.033098}
\begin{align}
    S_{l_s,\text{ }l_i} = \left|\int_0^{2\pi}\int_0^{2\pi}d\phi_sd\phi_iW(\phi_s,\phi_i) e^{il_s\phi_s} e^{il_i\phi_i}\right|^2
\end{align}
where $W(\phi_s,\phi_i) = \iint dq_sd q_i q_sq_i\Phi(q_s, q_i, \phi_s,\phi_i) $. 

When the spatial walk-off of the pump is small enough, the system possesses rotational symmetry. As a result, the two-photon wavefunction only depends on $\phi_s - \phi_i$ and the total OAM is conserved throughout the downconversion process \cite{kulkarni2018angular}. If the pump carries a net OAM of $l_p$, only the diagonal terms $S_{l_s, l_p-l_s}$ that satisfy $l_i + l_s = l_p$ remain nonzero. When the spatial walk-off of the pump cannot be neglected, the system no longer maintains rotational symmetry and the total OAM is not conserved (i.e., $l_s + l_i \neq l_p$). In this case, the off-diagonal terms that do not satisfy $l_s + l_i = l_p$ start to grow as "sidebands" in the OAM spectrum as shown in Figs. S1 -- S4 (Supplement 1).

Figure \ref{fig:fig1} shows the far-field intensity distribution of the signal field in collinear and non-collinear configuration. Here, without the loss of generality, we assume that the pump field has a Gaussian transverse spatial profile described by 
\begin{align}
    \tilde{E_p}(q_{px},q_{py}, z) = e^{-w_p^2 (q_{px}+q_{py})^2/4}e^{i(k_{pz}+ q_{px}\text{tan}\rho) z}
\end{align}
where $q_{px},q_{py}$ are the $x$- and $y$-components of the transverse wavevector, $w_p$ is the waist of the pump beam, $k_{pz}$ is the wavevector along the $z$-direction, and $\rho$ is the walk-off angle in the $xz$-plane. The magnitude of the walk-off angle describes the angular difference between the wavevector and the Poynting vector. In this study, we consider the SPDC process in a type-I $\beta$-BBO pumped by a continuous-wave laser beam with a wavelength of 355 nm. The length of the crystal $L$ and the waist of the pump field $w_p$ are 3 mm and 200 $\mu$m respectively. At collinear phase matching, the angle between the optical axis and the direction of pump propagation is 32.914$^\circ$ according to the Sellmeier relations of BBO \cite{eimerl1987optical}. Figures \ref{fig:fig1}a and \ref{fig:fig1}c present the intensity distribution of the downconverted signal in the far-field at collinear and non-collinear phase matching when no spatial walk-off effect is considered ($\rho = 0$). As expected, the generated intensity patterns preserve perfect rotational symmetry. In contrast, the broken rotational symmetry of the far-field intensity profiles  becomes an observable effect in Figs. \ref{fig:fig1}b (collinear) and \ref{fig:fig1}d (non-collinear) when the pump has a walk-off angle of $\rho = 0.052 $ ($\sim3^\circ$). In both cases, the central brightness of the downconverted signal field shifts along the $x$-axis, showing a preference for the direction of the Poynting vector.

We define an infidelity metric, $f_{\text{leak}}$, to characterize the degree at which OAM conservation is violated in the presence of the pump walk-off:
\begin{align}
    f_{\text{leak}} = 1 - \sum_{l_s = -\infty}^{\infty} S_{l_s, \text{ }l_{\text{tot}}-l_s }
\end{align}
where $l_{\text{tot}} = l_p$ is the total OAM of the system carried by the pump beam prior to the downconversion process. Essentially, $f_{\text{leak}}$ can be considered as a deviation in state fidelity from the perfect anti-correlated state $\ket{\psi} = \sum_l\ket{l}_s\ket{-l}_i$. For type-I SPDC without pump walk-off, the total OAM is conserved. Therefore, for a Gaussian pump ($l_p=l_{\text{tot}}=0$), the probability $S_{l_s,\text{ }-l_s}$ should sum up to unity for all possible values of $l_s$, namely, $f_{\text{leak}} = 1-\sum_{l_s} S_{l_s,\text{ }-l_s} = 0$. This indicates that the OAM of the signal and the OAM of the idler are perfectly anti-correlated. The OAM correlation matrix of the downconverted two-photon state only has non-zero anti-diagonal terms. 

When the pump walk-off becomes non-negligible, the phase mismatch $\Delta k$ in the two-photon wavefunction (Eq. \ref{eq:two-photon}) becomes 
\begin{align}
    \Delta k = k_{pz} - k_{sz} - k_{iz} + (q_s \text{cos}\phi_s + q_i \text{cos}\phi_i)\text{tan}\rho \label{eq:mismatch_walkoff}
\end{align}
where $k_{pz} = \sqrt{k_p^2 - (\mathbf{q}_s+\mathbf{q}_i)^2}$, $k_{sz} = \sqrt{k_s^2 - {q}_s^2}$, and $k_{iz} = \sqrt{k_i^2 - {q}_i^2}$ are the $z$-components of all participating wavevectors. With the extra walk-off term in Eq. \ref{eq:mismatch_walkoff}, the new two-photon wavefunction no longer depends solely on $\phi_s-\phi_i$. This leads to the appearance of non-zero cross-talk terms in the OAM correlation matrix because the total OAM is no longer conserved. The probability of observing a total OAM that is different from the initial pump OAM is greater than 0, which is equivalent to $f_{\text{leak}} > 0$. In this sense, the probability of conserving the total OAM "leaks" into other channels where the total OAM changes by at least one unit.

From Eq. \ref{eq:hamilt}, Eq. \ref{eq:two-photon} and Eq. \ref{eq:mismatch_walkoff}, we see that the walk-off effect depends on the propagation distance inside the nonlinear crystal and the focusing of the pump beam. To conveniently describe this dependence, we define a dimensionless focusing parameter $\sqrt{L/z_R}$ where $z_R$ is the Rayleigh range of the pump beam. Figure \ref{fig:fig2}a demonstrates the relation between the violation of total OAM and the focusing parameter for multiple pump walk-off angles ranging from 0.017 ($\sim1^\circ$) to 0.087 ($\sim5^\circ$) in collinear SPDC geometry. For a longer nonlinear crystal or a more focused pump, the generated OAM-entangled state becomes more susceptible to the spatial walk-off effect. On the other hand, we also notice that all curves asymptotically approach $f_{\text{leak}}=0$ when the crystal is sufficiently thin compared to the Rayleigh range of the pump beam. This behavior agrees well with the commonly used model for thin crystal approximation where pump walk-off is safely neglected \cite{PhysRevA.103.063508}. Figure \ref{fig:fig2}b shows the dependence $f_{\text{leak}}$ on  $\sqrt{L/z_R}$ for SPDC with non-collinear phase matching ($\theta = 39.935^\circ$). The non-collinear geometry shares the same asymptotic behavior when $\sqrt{L/z_R}$ approaches 0 as the collinear phase matching, but it appears to be less sensitive to the change of the focusing parameter and to the increase of the walk-off angle. Phase matching in the non-collinear case typically selects emission around a ring with a more constrained geometry. The transverse wavevectors of the signal-idler pair are strongly anti-correlated and confined in the proximity of a large transverse wavevector $q_0$ so that the projection of their transverse wavevectors on the walk-off direction varies more weakly than in the collinear case. Consequently, a significant part of the walk-off phase can behave like a nearly constant phase offset over the dominant emission region, and this nearly constant phase does not effectively mix OAM since only azimuthal variation does.

To better show the numerical dependence of the total OAM change on the walk-off angle, we plug Eq. \ref{eq:mismatch_walkoff} into the two-photon wavefunction in Eq. \ref{eq:two-photon} and apply the Jacobi-Anger expansion \cite{abramowitz1948handbook} of the walk-off term
\begin{align}
    e^{iq_{j}zcos\phi_j\text{tan}\rho} = \sum_{m= - \infty}^{\infty} i^m J_m(q_{j}z\text{tan}\rho) e^{im\phi_j} \label{eq:expand}
\end{align}
where $j = s,i$ is the subscript for signal and idler, $J_m(x)$ is the Bessel function of the first kind. The appearance of the term $e^{im\phi_j}$ in Eq. \ref{eq:expand} effectively couples the two OAM modes of the signal (idler) that differ by $m\hbar$. As a result, the total OAM distribution starts to have sidebands that satisfy $\Delta l_{\text{tot}} =\Delta(l_s + l_i) \neq 0 $ in addition to the central maximum that corresponds to $\Delta l_{\text{tot}} =0$.
Figures \ref{fig:fig3}a--c present the distribution of the total OAM for three different walk-off angles ($1^\circ$, $3^\circ$, and $5^\circ$) under the collinear phase matching condition. We notice that the sidebands of the total OAM distribution grow significantly as the walk-off angle becomes larger. Furthermore, it is usually safe to consider only the lowest order change in total OAM (i.e. $\Delta l_{\text{tot}} = \pm1$) for realistic pump walk-off effects in SPDC processes since higher-order changes (i.e. $\left|\Delta l_{\text{tot}}\right| >1$) are orders-of-magnitude smaller.

When the pump walk-off is small enough, that is, $q_j L \text{tan}\rho  \ll 1$ or $L \text{tan}\rho / w_p\ll 1$ in an experiment, the Bessel functions in Eq. \ref{eq:expand} can be approximated by their asymptotic forms \cite{abramowitz1948handbook}. Keeping only the two lowest orders, we have $J_0(x) \approx 1 - x^2/4 + ...$ and $J_{\pm1}(x) \approx \pm x/2 + ...$ for $x\ll 1$. From this approximation, a useful walk-off-dependent scaling law can be derived for the violation of total OAM conservation: the probability amplitude $S_{l_s, l_p + n - l_s}$ of the $n$-th order change in total OAM ($\Delta l_{\text{tot}} = n$) should scale with $(\text{tan}\rho)^{2|n|}$. For a small walk-off angle,the result can be further simplified to $S_{l_s, l_p + n - l_s} \propto \rho^{2|n|}$ since $\text{tan}\rho \approx \rho$. Figure \ref{fig:fig3}d demonstrates this scaling law in the small walk-off limit where the SPDC is driven by a Gaussian pump with $w_p = 0.5$ mm in a nonlinear crystal with a length of $L = 1$ mm. The colored solid lines (purple, maroon, and orange) are the polynomial fits of $(\text{tan}\rho)^2$, $(\text{tan}\rho)^4$ and $(\text{tan}\rho)^6$ for $l_{\text{tot}}=\pm1$, $\pm2$, and $\pm3$ respectively. The probability amplitudes of the total OAM change calculated from direct OAM projection are represented by shaped dots (circle, square, and triangle for $l_{\text{tot}}=\pm1$, $\pm2$, and $\pm3$). For most type-I SPDC crystals, the pump walk-off angle usually lies in the range from 1$^\circ$ to 5$^\circ$ \cite{yao1992accurate, feng2024pre}. The good agreement between the shaped dots and their corresponding line fits shown in Fig. \ref{fig:fig3} within this angle range justifies the use of the aforementioned scaling law in many realistic experimental settings.

\begin{figure}[hbt!]
\centering
\hspace*{-0.5cm}
\includegraphics[width=0.4\textwidth]{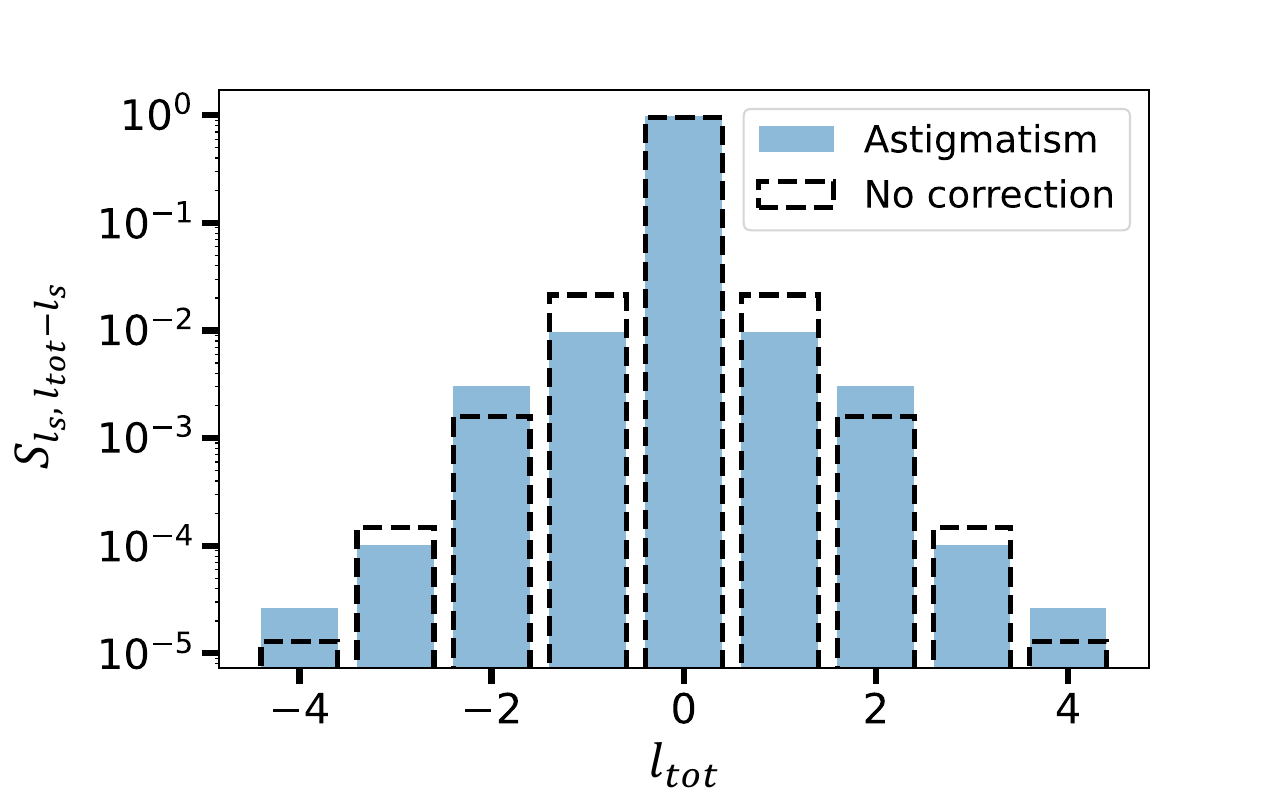}
\caption{Total OAM distribution of the downconverted signal-idler pair corrected by pump astigmatism. The downconverted photon pair is generated in a 3-mm-long type-I BBO with an estimated pump walk-off of $\rho \approx  4.6^\circ$ determined from $n_e \text{tan} \rho = -\partial n_e/\partial \theta$ \cite{eimerl1987optical}. The applied astigmatism suppresses the coupling of total OAM to odd-order azimuthal harmonics and enhances even-order changes due to its $\text{cos}(2\phi)$ dependence.}
\label{fig:fig4}
\end{figure}

Lastly, OAM mode engineering can be crucial to controlling the output quantum state of downconverted signal and idler photon for practical experimental purposes. With our pump walk-off model, we hereby propose a feasible method based on intrinsic wavefront engineering to accurately adjust the total OAM change by introducing controllable asymmetric features such as astigmatism and/or epllipticity to the spatial profile of the pump. For example, Fig. \ref{fig:fig4} compares the total OAM distribution with the ``optimal pump astigmatism correction" (Supplement 1) to an uncorrected one. The applied astigmatism reduces the rotational asymmetry of the system due to spatial walk-off along the x-axis and thus reduces the infidelity $f_{leak}$into the lowest-order sidebands ($\Delta l_{\text{tot} = \pm1}$), but we should note that the application of pure quadratic astigmatism suppresses only odd-order OAM sidebands but enhances even-order ones since standard Seidel astigmatism has a phase dependence of $\text{cos}(2\phi)$, which only leads to even-order azimuthal harmonics in Jacobi-Anger expansion. 

In conclusion, this Letter has demonstrated the effect of the spatial walk-off of the pump on the quantum state of the OAM-entangled photon pair produced by type-I SPDC. Our results have demonstrated a numerical relationship between the violation of total OAM conservation and the spatial walk-off angle. We have also proposed a scaling law for the total OAM change in the small pump walk-off limit. Our study provides useful guiding principles for the engineering of OAM-entangled photon pairs under realistic experimental conditions.

\begin{backmatter}

\bmsection{Acknowledgments}
We acknowledge the support from the following fundings: Department of Energy - Office of Science (FWP 762954); Canada First Research Excellence Fund; Canada Excellence Research Chairs, Government of Canada; Natural Sciences and Engineering Research Council of Canada.

\bmsection{Disclosures} The authors declare no conflicts of interest.

\bmsection{Data availability} Data underlying the results presented in this paper are not publicly available at this time but may be obtained from the authors upon reasonable request.

\bmsection{Supplemental document} See Supplement 1 for supporting content.

\end{backmatter}

\bibliography{main}

\bibliographyfullrefs{main}

\clearpage 

\onecolumn 

\begin{center}
    \textbf{\Large Full References}
\end{center}
\begin{enumerate}
    \item Y. Yan, G. Xie, M. P. Lavery, H. Huang, N. Ahmed, C. Bao, Y. Ren, Y. Cao, L. Li, Z. Zhao et al., “High-capacity millimetre-wave communications with orbital angular momentum multiplexing,” Nature Communications 5, 4876 (2014).
    \item A. E. Willner, Y. Ren, G. Xie, Y. Yan, L. Li, Z. Zhao, J. Wang, M. Tur, A. F. Molisch, and S. Ashrafi, “Recent advances in high-capacity free-space optical and radio-frequency communications using orbital angular momentum multiplexing,” Philosophical Transactions of the Royal Society A: Mathematical, Physical and Engineering Sciences 375, 20150439 (2017).
    \item Y. Xu, S. Choudhary, and R. W. Boyd, “Stimulated emission tomography for efficient characterization of spatial entanglement,” Physical Review Research 6, L042047 (2024).
    \item Z. Wang, R. Malaney, and B. Burnett, “Satellite-to-earth quantum key distribution via orbital angular momentum,” Physical Review Applied 14, 064031 (2020).
    \item D. Halevi, B. Lubotzky, K. Sulimany, E. G. Bowes, J. A. Hollingsworth, Y. Bromberg, and R. Rapaport, “High-dimensional quantum key distribution using orbital angular momentum of single photons from a colloidal quantum dot at room temperature,” Optica Quantum 2, 351–357 (2024).
    \item A. E. Willner, K. Pang, H. Song, K. Zou, and H. Zhou, “Orbital angular momentum of light for communications,” Applied Physics Reviews 8 (2021).
    \item M. Erhard, M. Krenn, and A. Zeilinger, “Advances in high-dimensional quantum entanglement,” Nature Reviews Physics 2, 365–381 (2020).
    \item L. Kopf, R. Barros, S. Prabhakar, E. Giese, and R. Fickler, “Conservation of angular momentum on a single-photon level,” Physical Review Letters 134, 203601 (2025).
    \item A. Mair, A. Vaziri, G. Weihs, and A. Zeilinger, “Entanglement of the orbital angular momentum states of photons,” Nature 412, 313–316 (2001).
    \item N. Quesada and J. Sipe, “Time-ordering effects in the generation of entangled photons using nonlinear optical processes,” Physical Review Letters 114, 093903 (2015).
    \item A. D’Errico, F. Hufnagel, F. Miatto, M. Rezaee, and E. Karimi, “Full-mode characterization of correlated photon pairs generated in spontaneous downconversion,” Optics Letters 46, 2388–2391 (2021).
    \item G. Kulkarni, J. Rioux, B. Braverman, M. V. Chekhova, and R. W. Boyd, “Classical model of spontaneous parametric down-conversion,” Physical Review Research 4, 033098 (2022).
    \item J. P. Torres, A. Alexandrescu, and L. Torner, “Quantum spiral bandwidth of entangled two-photon states,” Physical Review A 68, 050301 (2003).
    \item J. P. Torres, G. Molina-Terriza, and L. Torner, “Spatial entanglement and optimal single-mode coupling in photon pair generation,” Journal of Optics B: Quantum and Semiclassical Optics 7, 235–239 (2005).
    \item B. Kanseri and P. Sharma, “Effect of partially coherent pump on the spatial and spectral profiles of down-converted photons,” Journal of the Optical Society of America B 37, 505–512 (2020).
    \item Y. S. Lee, M. Xie, R. Tannous, and T. Jennewein, “Sagnac-type entangled photon source using only conventional polarization optics,” Quantum Science and Technology 6, 025004 (2021).
    \item S. Oh and T. Jennewein, “Polarization entanglement with highly nondegenerate photon pairs enhanced by an effective walk-off-compensation method,” Physical Review A 110, 063515 (2024).
    \item H.-J. Wu, B.-S. Yu, J.-Q. Jiang, C.-Y. Li, C. Rosales-Guzm\'an, S.-L. Liu, Z.-H. Zhu, and B.-S. Shi, “Violation of the optical subsystem's total angular momentum conservation due to spin-orbit coupling in multimode nonlinear optics,” Physical Review Letters 130, 153803 (2023).
    \item G. Kulkarni, L. Taneja, S. Aarav, and A. K. Jha, “Angular schmidt spectrum of entangled photons: derivation of an exact formula and experimental characterization for noncollinear phase matching,” Physical Review A 97, 063846 (2018).
    \item D. Eimerl, L. Davis, S. Velsko, E. Graham, and A. Zalkin, “Optical, mechanical, and thermal properties of barium borate,” Journal of applied physics 62, 1968–1983 (1987).
    \item B. Baghdasaryan, F. Steinlechner, and S. Fritzsche, "Justifying the thin-crystal approximation in spontaneous parametric down-conversion for collinear phase matching," Physical Review A, 103(6), 063508 (2021).
    \item M. Abramowitz and I. A. Stegun, Handbook of mathematical functions with formulas, graphs, and mathematical tables, vol. 55 (US Government printing office, 1948).
    \item J. Yao, W. Sheng, and W. Shi, “Accurate calculation of the optimum phase-matching parameters in three-wave interactions with biaxial nonlinear-optical crystals,” Journal of the Optical Society of America B 9, 891–902 (1992).
    \item T. Feng, R. Fan, R. Fan, C. Wang, C. Huang, A. Li, H. Zhang, Y. Cao, and L. Li, “Pre- and post-compensation to suppress birefringent walk-off effects of entangled photons,” Optics Express 32, 40283–40292 (2024).
\end{enumerate}

\end{document}